# Evaluation of onboard sensors for track geometry monitoring against conventional track recording measurements

Hengcheng Zhang [a], Zhan Yie Chin [a]*, Pietro Borghesani [a], James Pitt [c], Michael E. Cholette [b]

a: School of Mechanical and Manufacturing Engineering, UNSW Sydney, Australia
b: Science and Engineering Faculty, Queensland University of Technology, Brisbane Queensland, Australia
c: Queensland Rail Limited

* Corresponding author: jacky.chin@unsw.edu.au

**Abstract**

The main objective of this paper is to assess the estimation of track condition parameters using onboard micro-electro-mechanical-system (MEMS) accelerometers. A prototype of an onboard data acquisition system was designed and installed on a track recording car (TRC) and a measurement campaign was conducted on an extensive portion of the Brisbane Suburban railway network. Comparison of the accelerometer-based results vs TRC recordings have shown that accelerometers installed on the bogie are the best compromise between proximity to the source and insensitivity to impulsive noise. It was found that two vertical bogie accelerometers (left and right) provide a good quantitative estimate of vertical alignment and that strong correlations with TRC measurements exist for lateral MEMS accelerometer measurements (horizontal alignment). These findings suggest that two bogie MEMS accelerometers with vertical and lateral measurement axes are effective in estimating geographical distributions of vertical/horizontal alignment.

## 1. Introduction

To ensure safe operation of trains, track geometry must be maintained to within acceptable tolerances prescribed by a Registered Professional Engineer of Queensland (RPEQ). Track degradation can be divided in two main classes of issues: macroscopic geometry changes and localised defects. The former of which are the subject of this paper. Macroscopic geometry changes are quantified using parameters such as vertical and horizontal alignment, superelevation, gauge, and twist and compared to standards and/or company thresholds [1]. Traditionally, track geometry parameters are measured periodically using a versine system or a track recording car (TRC) [2]. These two measurement systems are considered a reliable reference for maintenance decision making, but both rely on scarce resources, especially when dealing with the allocation of a few TRCs over a vast network.

With the availability of cheaper sensors and data acquisition (DAQ) systems, TRC-based track monitoring can be complemented by track-monitoring systems installed on in-service vehicles. While not mature enough to replace TRCs, these systems are highly promising to improve the efficiency of the TRC allocation across the network. Onboard measurements obtained from in-service trains can provide frequent assessments of the geographical distribution of track deterioration, which would in turn guide the prioritisation of TRC scheduling for the sections of the network at higher risk of exceeding degradation thresholds. A first generation of onboard monitoring systems for in-service vehicle is known as the unattended geometry measurement systems (UGMS) [3]. UGMSs measure the full geometry of tracks like the TRC by using a combination of inertial sensors and optical/laser sensors. The advantages of UGMS are related to the accuracy of their estimates, but their drawbacks are the high cost of the system, hindering its wide deployment across a fleet of trains and the need to regularly clean the optical sensors. A more cost-effective approach is to use the inertial sensors only, without the help of optical sensors. This approach is based on the assumption that the direct measurement of track geometry is not necessary for detecting track degradation, and that track quality is in essence captured by its effect on the dynamics of axles, bogies, and car body. Examples of these more cost-effective on-board monitoring systems (with only accelerometers and gyroscopes) are available in literature [4]–[6].

Inertial sensors on the axle box are directly related to the track geometry if the deformation of wheels and wheel-rail connection are neglected [7]. In the vertical direction, the assumption of a purely algebraic relationship between axle-box measurements and track profile is very reasonable. The flexibility of the wheels in the frequency range of interest for geometry measurements is negligible, and the first resonances are significantly above this range. The only possible biases result from potential motion of the rail itself under load, and impulsive events due to localised track defects. The first are usually neglected as they also affect TRC measurements. The second are in principle separable from the geometry owing to their high-frequency content, but practically they tend to saturate sensor readings compromising the entire range of measurement, or conversely require high-g accelerometers which have a lower sensitivity. Sensors on the bogie are less affected by impulsive and broadband noise due to the vibration isolation provided by the primary suspensions, which decrease the maximum acceleration peaks from about 100g on the axle box to 10g on the bogie [8]. Weston et.al. [9], [10] indicated that the bogie acceleration was affected by geometric filtering, which bring the transfer function between axle and bogie to zero at twice the bogie wheelbase, therefore, gyroscopes were utilised to replace accelerometer, and which gave acceptable results. However, the geometric filtering effect should only exist on the centre of the bogie, which could be solved by putting accelerometers on corners of the bogie. Acceleration of the car body is further filtered by the secondary suspensions, resulting in less than 1g peak acceleration, but this could also filter out vibration components symptomatic of the track conditions.

Portable condition monitoring systems with accelerometers, gyroscopes and noise level meters were developed and put on a car body by Mori et.al. [11], [12], and the root-mean-square (RMS) was found to trend with track geometry measurements on a TRC. By analysing a dynamic model of train, Alfi et al. [13] numerically demonstrated the intuitive understanding that sensors on the axle box are more effective in capturing short wavelength track irregularities, bogie acceleration is effective for medium wavelength, and car body acceleration is effective for the detection of long wavelength track irregularities. The exact frequency range in which each of the three options is optimal depends on the combination of speed (that drives the excitation frequency range) and vehicle dynamic characteristics, resulting in the location of the resonances in the transfer function between the excitation and the measurements. In their paper, Alfi et al. therefore suggested to fuse these measurements by using dedicated weighting functions, optimally tuned for each specific case. A recently published paper [14] has further confirmed that RMS of bogie measurements correlates well with vertical irregularities of track, and that once this correlation is established, it can be used to estimate track condition based on bogie acceleration measurements. Of course, the authors concluded that these correlations are characteristics of the specific train-track system under investigation, and therefore only valid for a specific fleet of similar trains on a specific line.

The main objectives of this paper are therefore to: (i) assess the feasibility of using budget micro-electro-mechanical-system (MEMS) sensors rather than more expensive and usually higher-performance piezo-electric sensors, thus reducing significantly the overall onboard system cost and consequently easing their deployment across a vast fleet of trains, (ii) compare the effectiveness of sensors installed in different locations (axle, bogie, and car body) to infer key track condition parameters (e.g., vertical and horizontal alignment instead of the full geometry) in a speed range compatible with urban and suburban trains (~10-60 km/h), (iii) do so without any information on the train or any requirement in terms of historical data to obtain a correlation like that in [14], and (iv) provide some insights based on an observed phenomenon of track geometry measurements in the frequency domain.

To achieve these tasks, a prototype of an onboard data acquisition (DAQ) system was designed and installed on the track recording car (TRC), which includes a number of redundant MEMS and "reference" piezo accelerometers on the axle box, bogie and car body to compare their performance and identify the best layout. A measurement campaign was conducted by running this TRC in the Brisbane Suburban Area (BSA) network. A series of signal processing procedures were then developed and applied to the vibration data to extract estimates of vertical and horizontal alignment without historical data and/or models, and their results were compared with the corresponding TRC measurements. The results obtained independently using sensors on different positions (axle, bogie, and car body) were compared with each other to find the best sensor position, and in addition, the influence of noise for the budget MEMS sensors were studied.

The remainder of the paper is organised as follows. The experimental setup is given in Section 2, methodology for track condition estimation is provided in Section 3, and the analysis results are presented in Section 4. At last, discussions and conclusions are drawn in Section 5.

## 2. Experimental setup

The experiments were conducted on a TRC instrumented with an additional on-board acceleration measurement system. In order to identify and validate the best layout of sensors for the estimation of track condition parameters and compare the performance of different types of sensors, a redundant number of sensors were installed, including 8 MEMS accelerometers and 3 integrated-electronics-piezo-electric (IEPE) accelerometers. The MEMS sensors have been chosen because of their lower price and sufficient performance, while the IEPE sensors represent the traditional reference (best-performing) vibration measurements. The sensor layout on the TRC for this campaign is shown in Figure 1, and consisted of:

- Axle box: 2 MEMS sensors and 1 IEPE sensor
- Bogie: 3 MEMS sensors and 1 IEPE sensor
- Car body: 3 MEMS sensors and 1 IEPE sensor

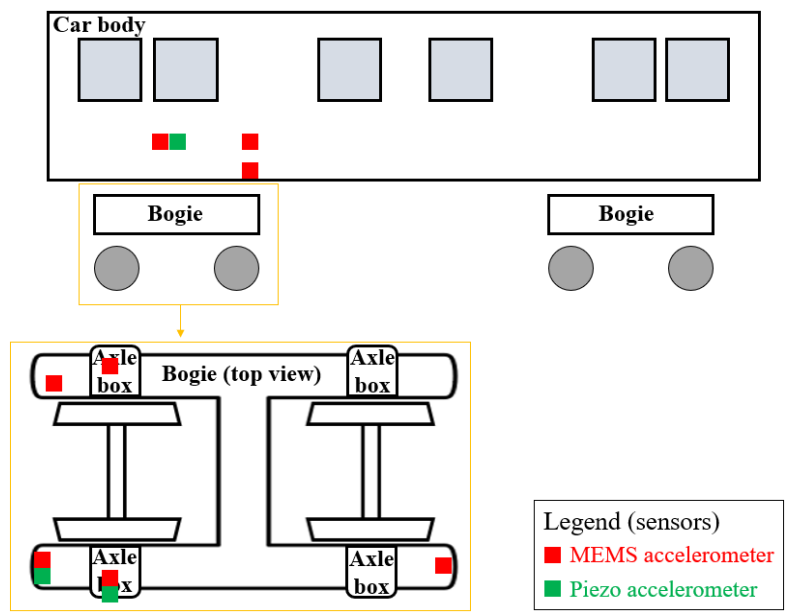

*Figure 1. Schematic diagram of the sensor layout.*

Each MEMS sensor has measurements in 3 directions (vertical, lateral, and longitudinal) integrated together, but not all the channels were acquired by the DAQ. For MEMS sensors on the axle box, acceleration in the vertical and lateral directions were collected. For the bogie, vertical and lateral measurements were taken from the two MEMS sensors at the front, while only vertical measurement was acquired from the rear one. For the 3 MEMS sensors on the car body, only vertical acceleration was used. The IEPE sensor can only measure acceleration in one direction, and they were all placed to measure acceleration in the vertical direction. The characteristics of the sensors are reported in Table 1, and they were acquired synchronously using a multi-channel DAQ system and a sample rate of 2,560 samples/s. Measurements were collected while running the TRC on the Brisbane Suburban Area railway network.

*Table 1. Sensor characteristics*

| Sensor type | Installation position | Range (g) | Noise level ($\mu g/\sqrt{Hz}$) |
|---|---|---|---|
| MEMS | Car body | 3 | 150 |
| MEMS | Bogie | 16 | 300 |
| MEMS | Axle box | 200 | 2700 |
| IEPE | Car body and bogie | 50 | 3 |
| IEPE | Axle box | 500 | 16 |

**3. Methodology for track condition estimation**

In this section, detailed procedures on how the vibration data was processed to obtain track condition estimates comparable to those measured by the TRC are provided. The analysis results and discussion are presented in the next section.

As illustrated in Figure 2, all vibration records were first down-sampled and merged so that they correspond to a TRC file representing a particular section of the track. Vibration of the test sensors were simultaneously recorded at the sampling rate of 2,560 Hz, and the data was saved to a file every 10 seconds. To save computing power and to compare with the TRC measurements, the vibration files corresponding to approximately the same time period of the TRC records were down sampled to a decimated value of $f_s$ = 256 Hz, and then attached together.

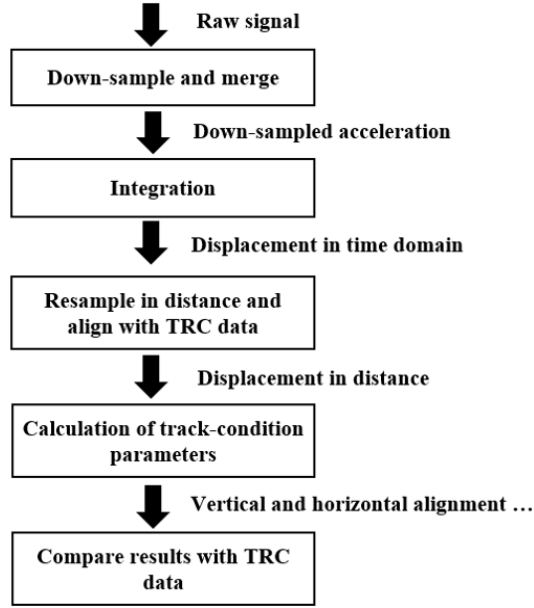

*Figure 2. Processing of vibration data.*

The down-sampled acceleration signals were then integrated twice with respect to time to obtain the corresponding displacement signals. This was done in the frequency domain, by dividing the discrete Fourier transform (DFT) of the signal by the frequency axis twice:

$$z[n] = \mathcal{F}^{-1}\left\{\frac{\mathcal{F}\{a_z[n]\}[k]}{(2\pi f[k])^2}\right\}[n] \tag{1}$$

where $a_z[n] = a_z(n/f_s)$ represents the sampled acceleration signal (with sampling rate $f_s$ samples per seconds), $z[n]$ the corresponding displacement signal (with $n = 0, ..., L-1$ samples) and $f[k]$ the frequency axis of the corresponding DFT $\mathcal{F}\{\ \}$. Proper high-pass filtering was also implemented to sufficiently remove low-frequency noise in the signal, which would otherwise be amplified by the integration process.

Although the vibration data files were chosen corresponding to the same period of the TRC measurements, they are not in sufficient alignment because their time stamps were not synchronised. Moreover, the TRC data was sampled uniformly in space (4 samples/m) while the vibration data was sampled uniformly in time (2560 samples/s), independently of the instantaneous speed of the vehicle. To mitigate this situation, the instantaneous speed of the vehicle is required for both resampling the vibration data at regular spatial intervals (i.e., obtain one sample every 250 mm of track) and for estimating the correspondence between certain instants with precise locations along the track in later steps of the processing procedure.

A Global Positioning System (GPS) module was not included in the onboard system for this test campaign. While GPS-based speed measurements were not available, the instantaneous vehicle speed could be estimated from the vibration records, thanks to the availability of two bogie vertical measurements on the same side of the bogie, one at the front and the other at the back as mentioned in Section 2. These measurements are expected to have a strong correlation, albeit with a time-varying time-delay. Given that this speed estimation step is not an essential part of the processing procedure and speed can be alternatively measured by a GPS module, the complete speed estimation procedure has been included in the Appendix. The authors would like to highlight that precise point-by-point comparison (i.e., with resolution of 250 mm) of the data is not necessarily valid due to the approximate realignment of the data using the speed estimated by the cross correlation of the front and back bogie measurements. Later sections will show that the results are in good correspondence when compared over a greater portion of the railway network in 100-m windows, which provides sufficient data for the rail industry to make maintenance decisions, such as the scheduling of a tamping machine. However, the authors would also like to emphasise that the speed estimation procedure (thus calculation of distance axis) is by no means intended to replace GPS measurements in industrial practice, but instead serves as a temporary contingency step for the absence of a GPS device in this particular experimental campaign. Future campaigns will include GPS data and aims to provide more accurate positioning for the track geometry estimates, which will potentially allow for analysis of more localised or shorter-wavelength defects. For an actual application on passenger/freight trains, the presence of a GPS in the onboard system will actually be necessary, since only on the TRC it was possible to take advantage of the TRC's GPS system.

Figure 3 demonstrates the alignment of the speed profile estimated from bogie measurements and that of the TRC records, using the data corresponding to the track portion from Eagle Junction to Bowen Hills as an example.

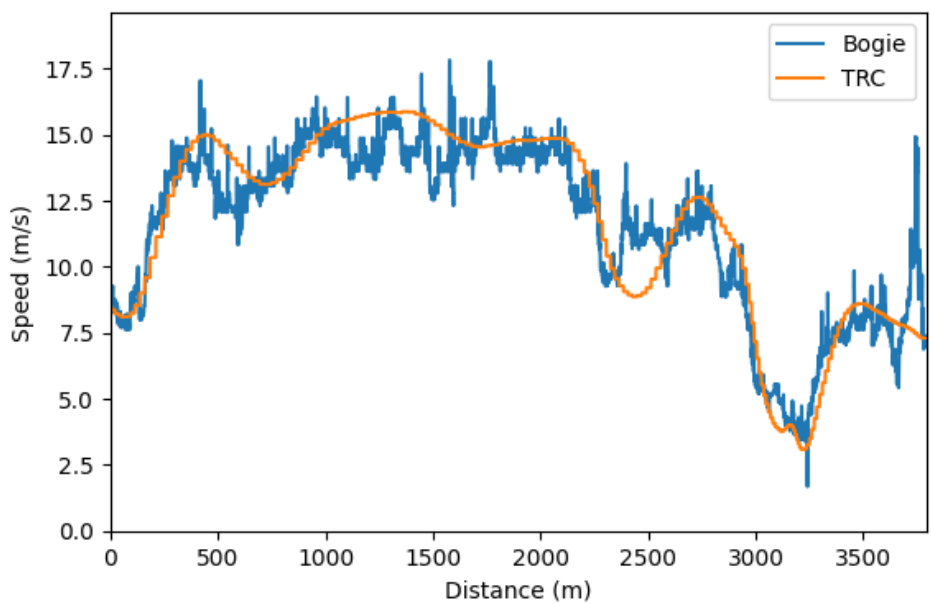

*Figure 3. Alignment of vibration-based (bogie acceleration) and TRC speed profiles.*

Using the estimated instantaneous vehicle speed, the mapping of track position vs time could then be obtained, which allows for the resampling of the vibration data from time domain to distance domain in 4 samples per meter, consistent with the TRC data. The distance axis along the track was built by integrating speed with respect to time:

$$x(t) = x_0 + \int speed(t)dt \quad (2)$$

The integral was performed as a discrete sum of estimated instantaneous speed multiplied by unit time:

$$\int speed(t)dt \approx \sum speed[n\delta t] \cdot \delta t \quad (3)$$

The unit time $\delta t$ is $1/f_s$ second, $speed[n\delta t]$ is the instantaneous speed at each point, and $x_0$ is the location at the beginning of each record, which was adjusted in this case by aligning the speed profiles shown in Figure 3.

Now that the displacement signal in time domain is in correspondence with the built distance axis $x_0$, a new axis $x_{new}$ with unit point distance of 0.25 meter was built, thus the displacement signal could be resampled accordingly.

The next step was to calculate the track-condition parameters, e.g., vertical alignment ($VA$) and horizontal alignment ($HA$), for a single rail from the resampled displacement signal by using analytical formulas of their definition:

$$VA_d(x) = z(x) - \frac{1}{2}\left(z\left(x - \frac{d}{2}\right) + z\left(x + \frac{d}{2}\right)\right) \quad (4)$$

where $d$ is the chord length, and $z(x)$ is the vertical displacement, and

$$HA_d(x) = y(x) - \frac{1}{2}\left(y\left(x - \frac{d}{2}\right) + y\left(x + \frac{d}{2}\right)\right) \quad (5)$$

where $y(x)$ is the lateral displacement.

Figure 4 shows a visual example of the calculation of vertical alignment $VA_d(x)$ from the vertical measurement $z(x)$ with chord length $d$, with an exaggerated track profile for ease of visibility and understanding. In a real scenario, the distance between the two ends of the chord is almost identical to its horizontal projection $d$ since the vertical change between $z(x - d/2)$ and $z(x + d/2)$ is order of magnitudes smaller (up to tens of mm) than the chord length (m). Therefore, in this paper, the two quantities are used interchangeably.

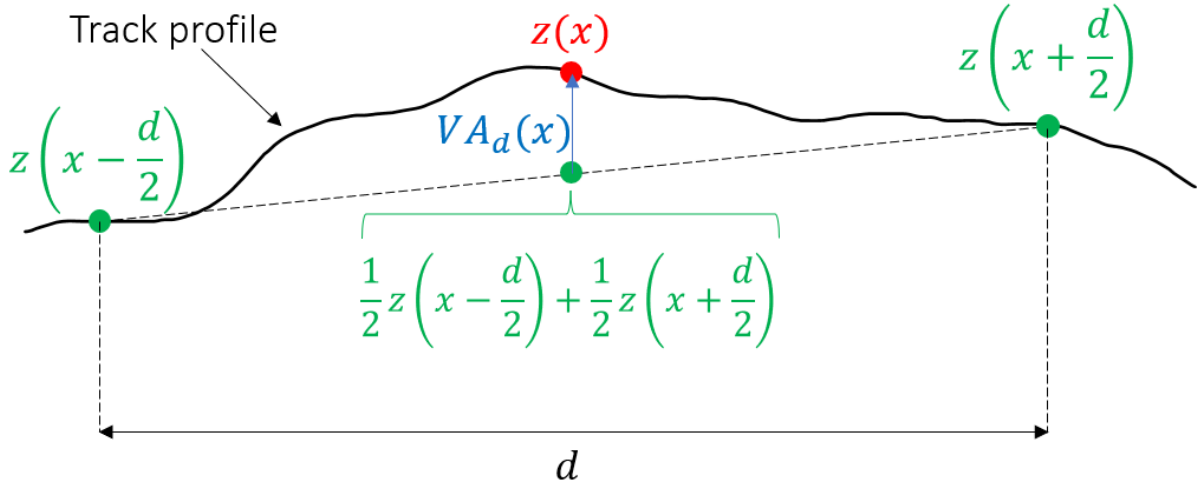

Figure 4: Visual example of the calculation of $VA_d(x)$.

In Figure 5, the parameter $VA_{10}$, which represents the vertical alignment calculated with chord length of 10 m, of the left rail obtained from vibration signals (resampled every 250 mm of track) in comparison with the corresponding TRC data from Eagle Junction to Bowen Hills is shown as an example (left column), with zooms (right column) for better illustration.

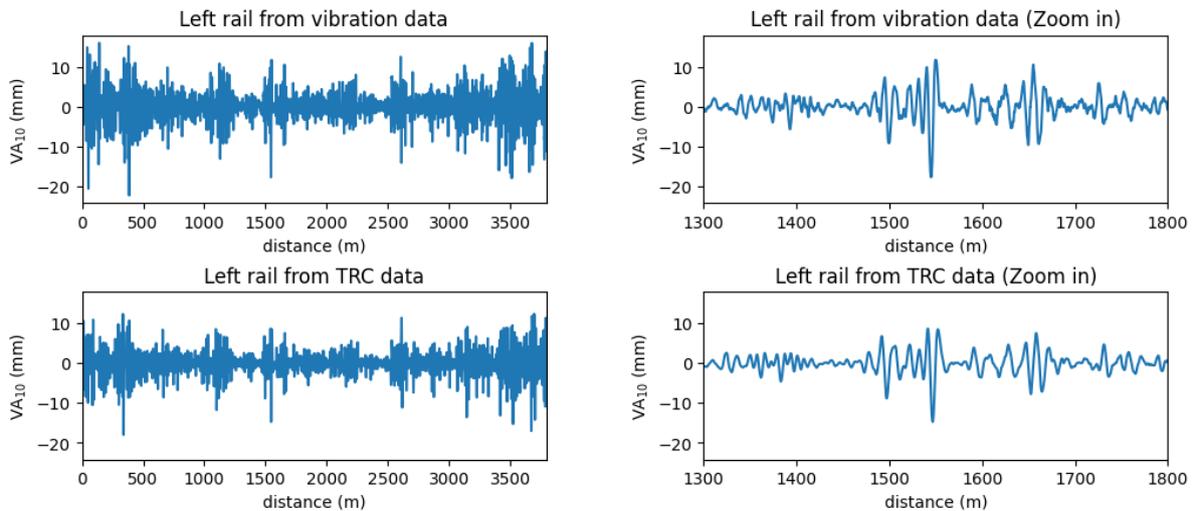

Figure 5. Example of $VA_{10}$ sampled every 0.25 m of track, vibration (top) vs TRC data (bottom).

In the last step, the track condition parameters estimated from vibration data and those from TRC were compared by their maximum values over a moving window of 100 m (coarse) or 20 m (fine). Their correlation was calculated, and they were plotted on a geographic map for better illustration.

## 4. Results

In this section, the results of the signal processing procedures stated in Section 3 are presented. This section first analyses results from different sensor locations and the sensor noise to validate the choices of sensor placement and sensor types. It is anticipated that the primary purpose of the on-board system will be to prioritise the (more detailed, more resource intensive) TRC inspections across the network. Therefore, a high-level comparison of track-condition parameters was considered the most practical way to estimate the capability of onboard sensors to detect key areas in the network that require attention. Rather than an exact match between TRC measurements and onboard sensor results, the comparison will focus on the capability of onboard sensors to pinpoint geographical areas with critical levels of track degradation. Therefore, the next subsections of the results will present geographical maps of track condition estimated using two bogie accelerometers and compare them to the corresponding results obtained from the TRC data. Later subsections will instead provide more in-depth analysis of the results for a selected track, which is used as an example to show more details, and to provide some insights via the frequency analysis of track geometry measurements.

### 4.1 Comparison of sensors on different location

In this section, the comparison of the results between three locations, bogie, axle box, and car body are presented, so as to support the final selection of the bogie sensor measurements to be used later sections. Using the Eagle Junction to Bowen Hills period as an example, the maximum $VA_{10}$ of every 100 meters are calculated, as shown in Figure 6.

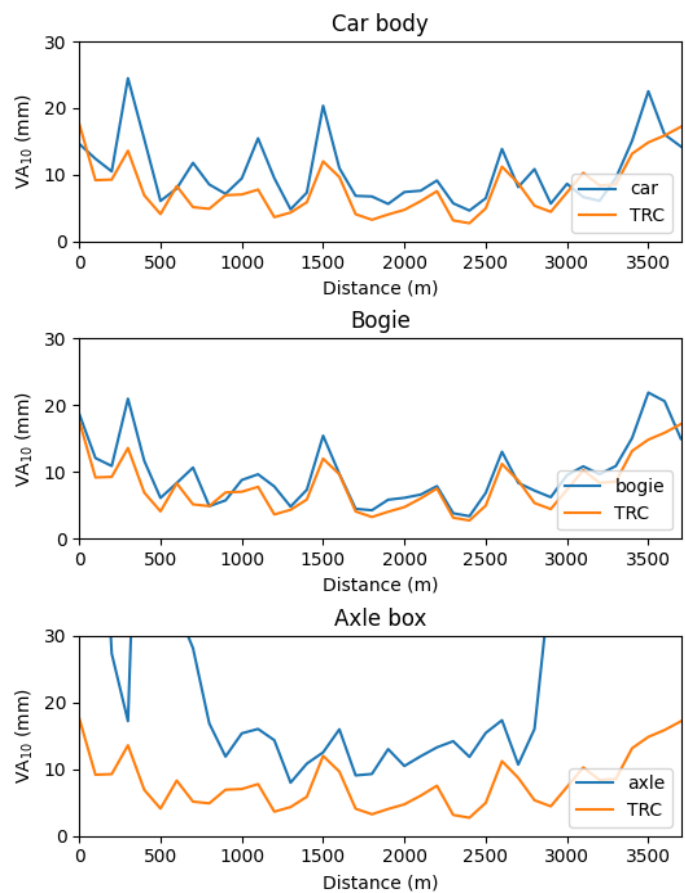

*Figure 6. Comparison of sensors on different locations: (top) car body, (middle) bogie, and (bottom) axle box.*

The results show that the car body measurements are less sensitive to specific events on left and right side compared to the bogie sensors, which is to be expected due to the mixing of left and right rail input in the vibration of the car body. The axle box measurements are instead negatively affected by strong impulsive events, resulting in extremely high-power events. This interpretation is further validated by the observation of full resolution $VA_{10}$ signals obtained from the axle box sensors, shown in Figure 7. Since it is not reasonable to expect $VA_{10}$ values of the magnitude shown in Figure 7 (top left), it is concluded that the extremely high values are due to impulsive noise which is integrated into unrealistically large displacements.

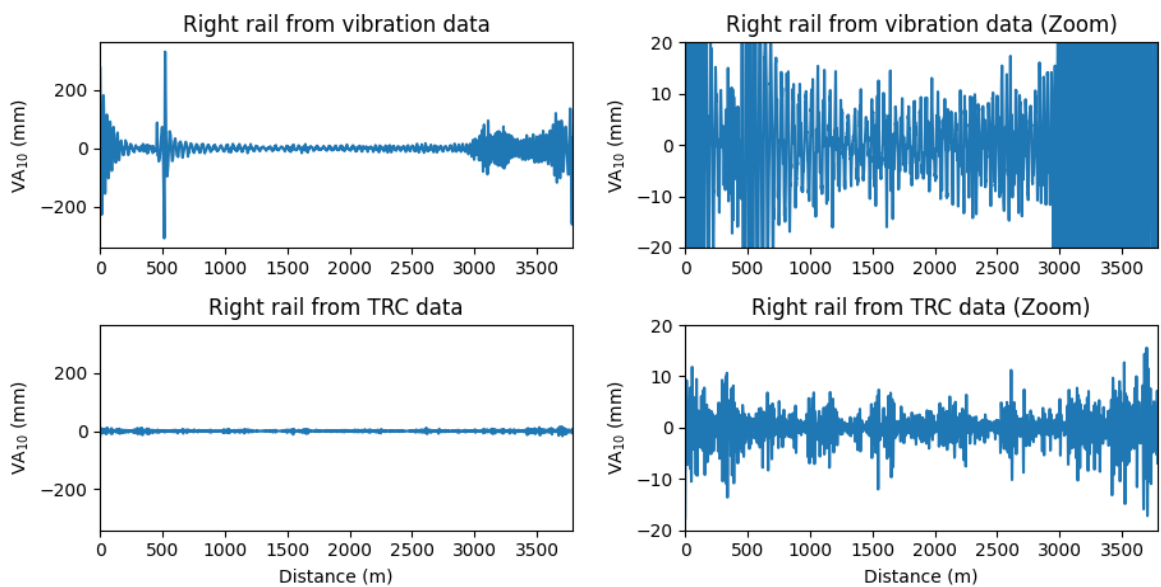

*Figure 7. Comparison of $VA_{10}$ measurements derived from axle box sensors and TRC data.*

The MEMS sensors are also compared with the high-performance IEPE sensors at the same positions. The power spectral density (PSD) of the sensors at different locations at about 8 m/s vehicle speed are presented in Figure 8, and the specified noise levels of the MEMS sensors are also plotted as reference. It shows that the MEMS sensors are in good agreement with the IEPE sensors under 100 Hz (corresponding to 7 cm wavelength in this case), which is sufficiently high. The bogie shows a 20-30 dB difference between the measurement and the theoretical noise floor of the sensor, compared to ~10 dB of carbody and axlebox sensors, confirming the suitable selection of bogie measurements in later sections. The car body measurements are particularly close to the theoretical noise level in high frequency (short wavelength) range, whereas the axle box measurements are particularly unperforming in the low frequency (long wavelength) range. These findings agree with the analytical model proposed in Ref. [13] and confirm that the bogie sensor location offers the best trade-off in terms of sensitivity to the actual track profile signal and noise (both higher the closer the sensor is to the track).

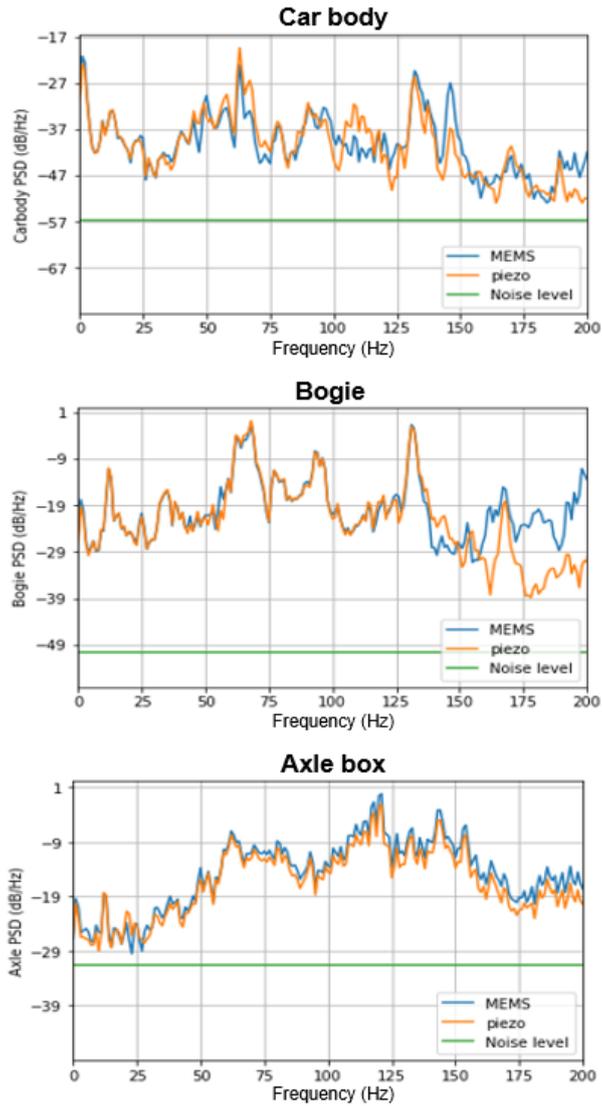

*Figure 8. Acceleration PSD for comparison of noise level of sensors: (top) car body, (middle) bogie, and (bottom) axle box.*

### 4.2 Overall geographical distribution of $VA_{10}$ (bogie measurements)

The following figures show the overall geographical distribution of $VA_{10}$ maxima over 100 m long portions of the whole track covered by this experimental campaign. Each diagram shows a comparison between TRC measurements (left diagram), and estimates based on bogie accelerometers (right diagram). The main areas of concern identified by the accelerometer estimates are very similar to the TRC.

The results of the left and right rail measurements during the southbound running of the TRC are presented in Figure 9 and Figure 10 respectively. Comparing them, it can be seen that the left and right rails have very similar trend. The $VA_{10}$ distribution of the left rail for northbound running is shown in Figure 11. It shares the same track from Roma Street to Sandgate with the southbound running, and their comparable results confirm the repeatability of this method. The Eagle Junction to Northgate period is missing because the TRC did not take a record there.

**Southbound Left $VA_{10}$, max over 100 m**

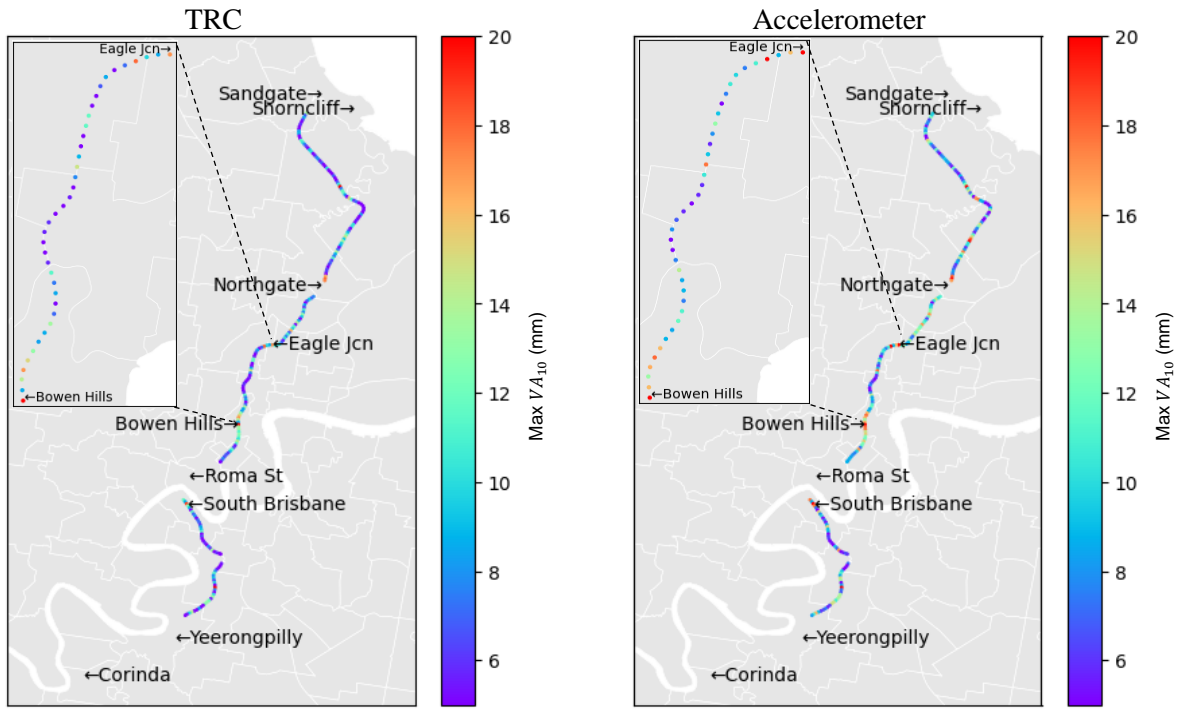

*Figure 9. Southbound Left $VA_{10}$, max over 100 m: (left) TRC data, (right) Bogie sensor.*

**Southbound Right $VA_{10}$, max over 100 m**

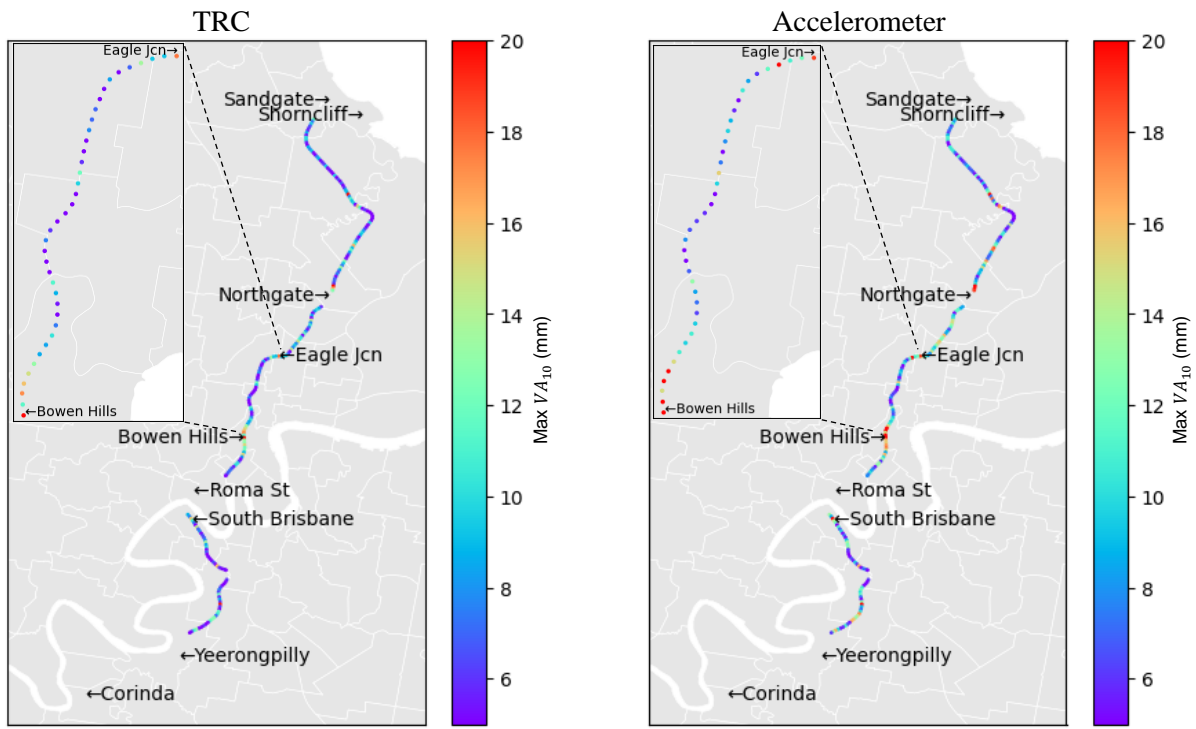

*Figure 10. Southbound Right $VA_{10}$, max over 100 m: (left) TRC data, (right) Bogie sensor.*

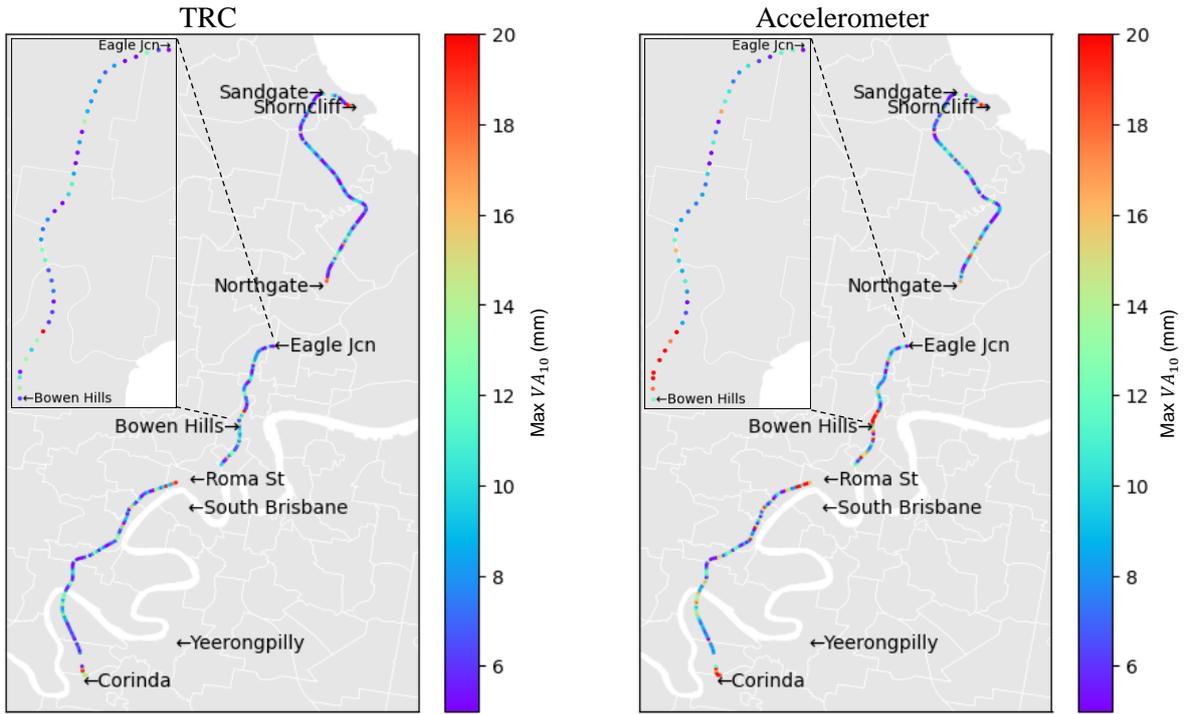

*Figure 11. Northbound Left $VA_{10}$, max over 100 m: (left) TRC data, (right) Bogie sensor.*

## 4.3 Overall geographical distribution of $HA_{10}$ (bogie measurements)

Similarly to the vertical geometry of the track represented by $VA_{10}$, the analysis of track lateral geometry by using $HA_{10}$ is illustrated in Figure 12 and Figure 13. Although the results derived from the vibration signals are obviously showing higher overall noise (higher maxima especially on "good" tracks), the key areas of concern are clearly identified nonetheless, with the highest $HA_{10}$ values located in the same geographical areas as for the TRC data. The overestimation is probably due to the fact that the measurement of bogie lateral acceleration is affected by the roll and yaw motion of the bogie, and not only associated with the track irregularity input. Comparing southbound and northbound runs in Figure 12 and Figure 13, a similar trend is observed, validating the repeatability of the results.

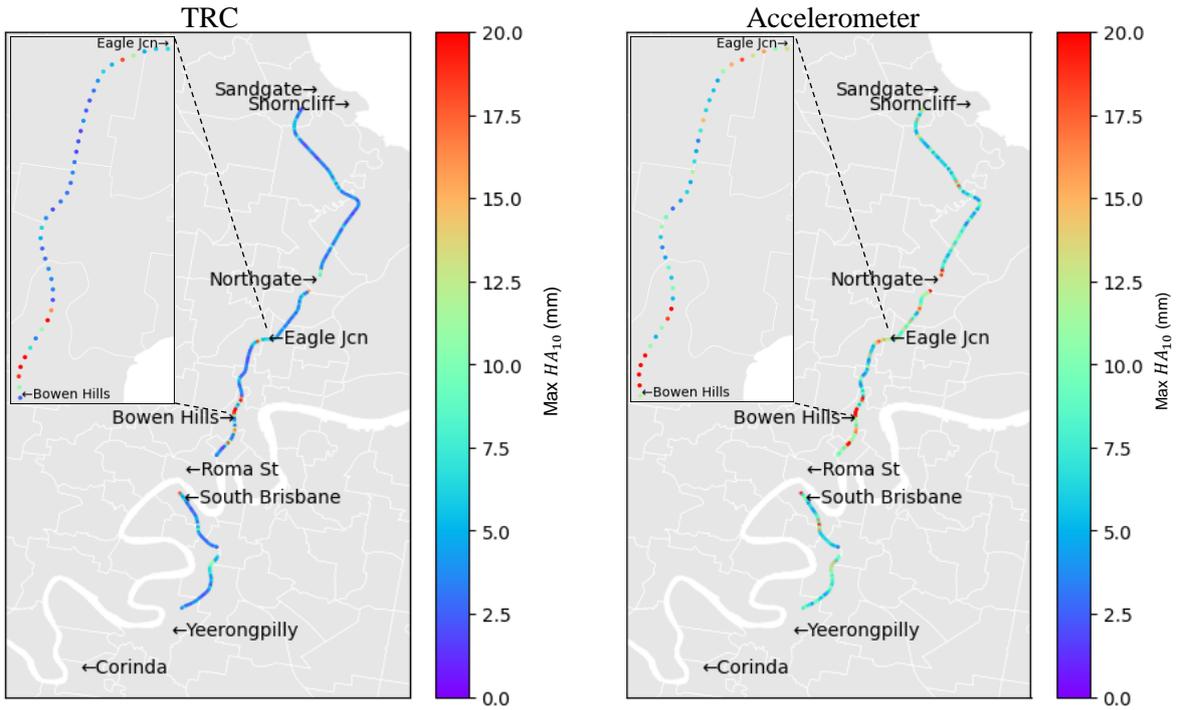

*Figure 12. Southbound Left $HA_{10}$, max over 100 m: (left) TRC data, (right) Bogie sensor.*

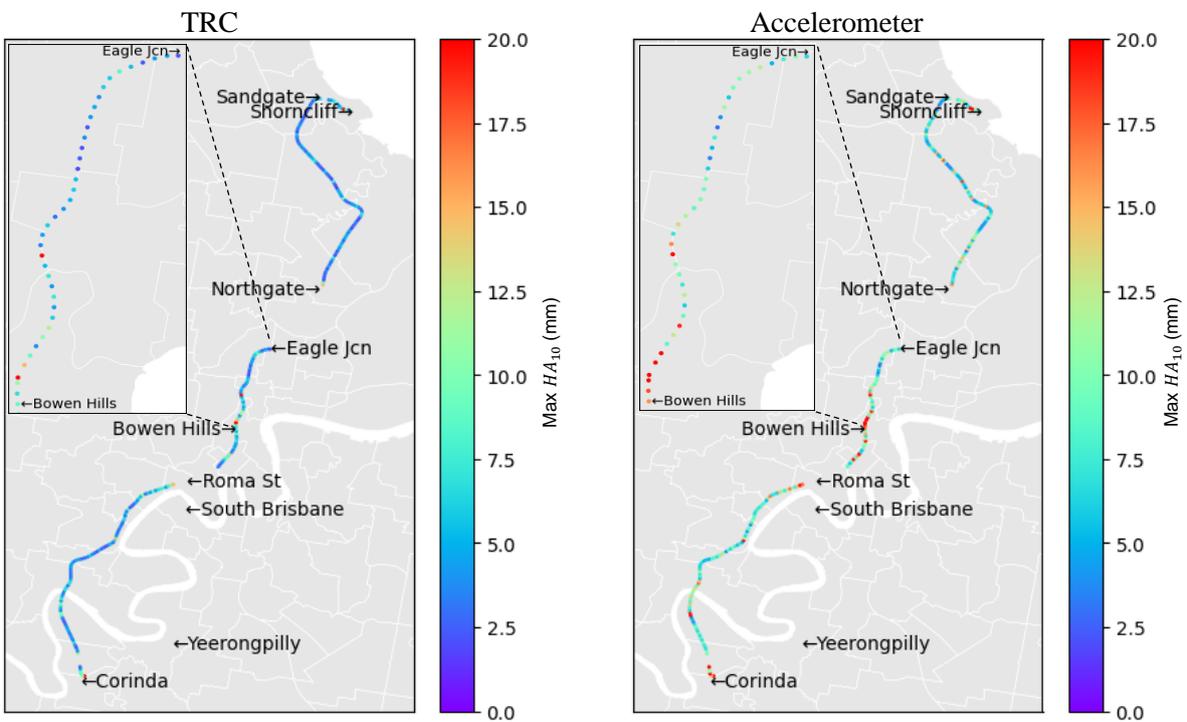

*Figure 13. Northbound Left $HA_{10}$, max over 100 m: (left) TRC data, (right) Bogie sensor.*

## 4.4 Detailed results on Eagle Junction to Bowen Hills (bogie measurements)

The overall results for the whole test campaign are shown in the previous sections, while in this section, detailed analysis on the period of Eagle Junction to Bowen Hills, which corresponds to the period representing the speed profiles in Figure 3, is presented as an example. The correlation between vibration-based estimates and the TRC measurements is also studied. It is important to note that the high-pass cutoff frequency of the acceleration signals was set at 0.3 Hz for $VA_{10}$ and $HA_{10}$, and 0.1 Hz for $VA_{35}$. Some perspective regarding the reasons for these choices, together with a frequency analysis of the measurements, will be discussed in the next section.

The results for the maximum value of $VA_{10}$ measurements every 100 meters are shown in Figure 14, and the correlation is presented on the right. It can be seen that the results agree well for both left and right rail, and their correlation coefficients are 72%-83% with offsets. It is to be noted that most of the correlation errors are likely due to an imperfect alignment between the vibration-based vs TRC data, which was adjusted but could not be fully corrected, given the accumulated small speed estimation errors over long distances. This alignment error is only relevant for this campaign due to the speed estimation based on bogie accelerometers, and these errors will be reduced in future campaigns by including a GPS device in the onboard track monitoring system for geographical alignment in future campaigns.

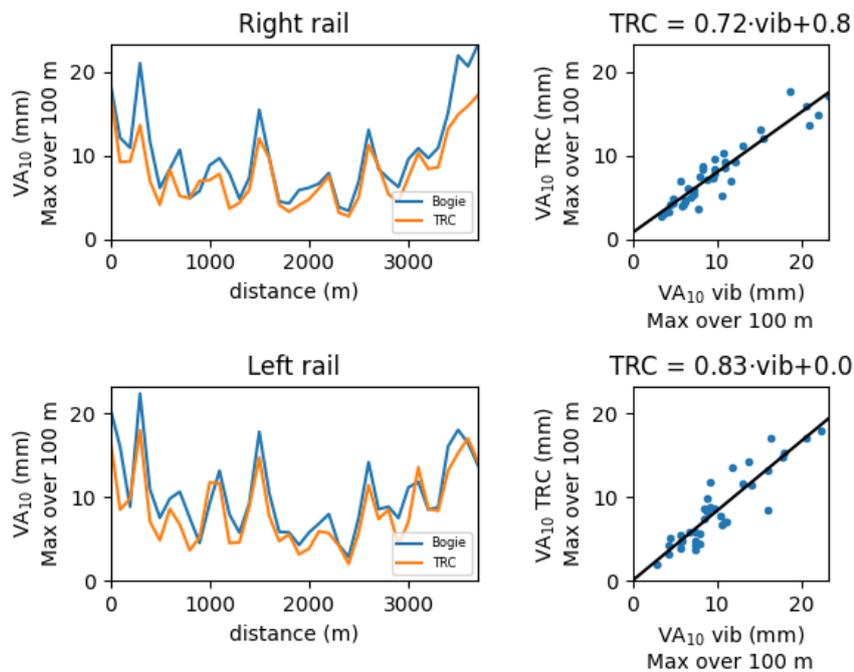

*Figure 14. Comparison of maximum $VA_{10}$ of every 100 meters.*

The same comparison was conducted for $VA_{10}$ maxima over a 20-meter moving window, and the results are similar as shown in Figure 15, with correlation coefficients further decreasing due to the higher impact of an imperfect data alignment on shorter windows.

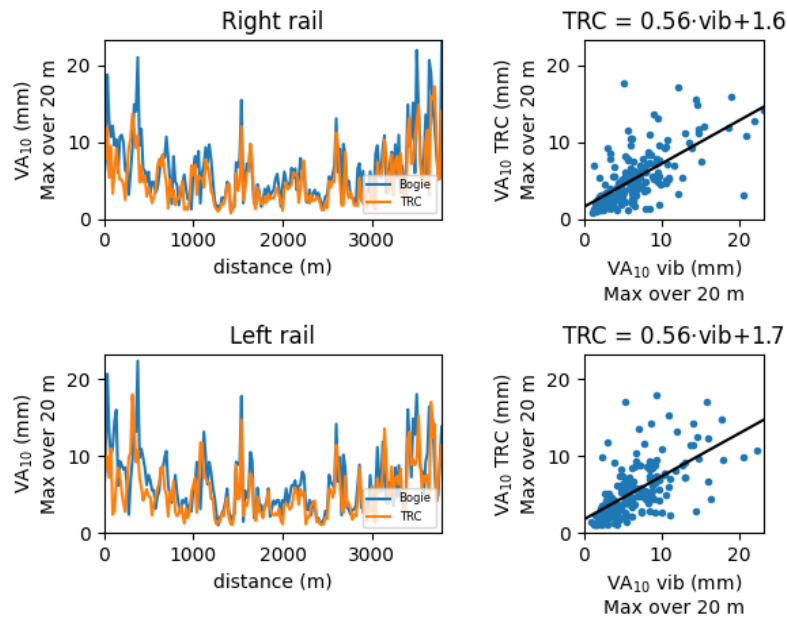

*Figure 15. Comparison of maximum $VA_{10}$ of every 20 meters.*

The maxima of another common track geometry parameter, i.e., $VA_{35}$, calculated over a 100 m window are also given in Figure 16, which show even better correlations, with the longer chord (35 m vs 10 m) reducing the impact of misalignment, since it acts as a low-pass filter.

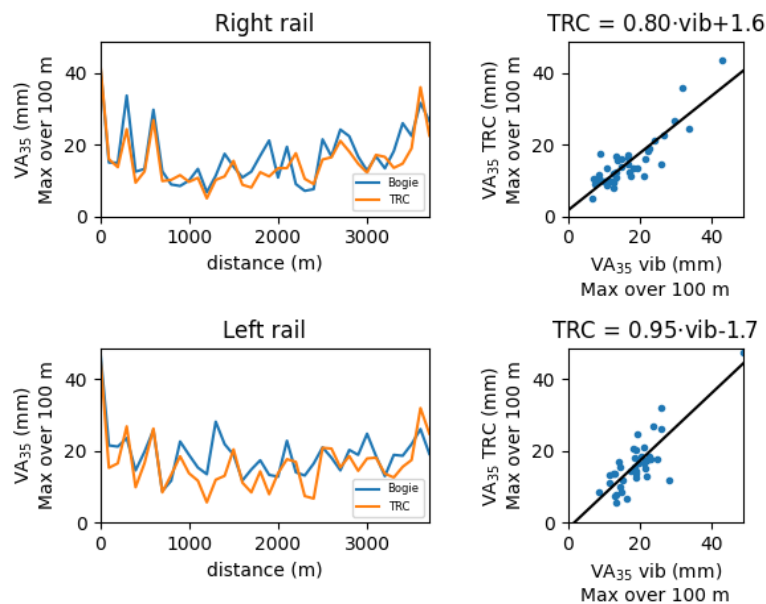

*Figure 16. Comparison of maximum $VA_{35}$ of every 100 meters.*

The same procedures are performed on horizontal alignment ($HA$). The direct comparison of $HA_{10}$ is shown in Figure 17, and the maximum values of every 100 meters are illustrated in Figure 18. The results from vibration show some obvious over-estimations, probably due to horizontal motion of the bogie triggered by either roll-related events (cross-talk between vertical and horizontal dynamics of the bogie) or impulsive events/vehicle dynamics of the bogie (horizontal/yaw vibrations). Nonetheless, it is interesting to observe that the two maximum values overall (over the entire track) seem to match, possibly suggesting that some correlations could be established.

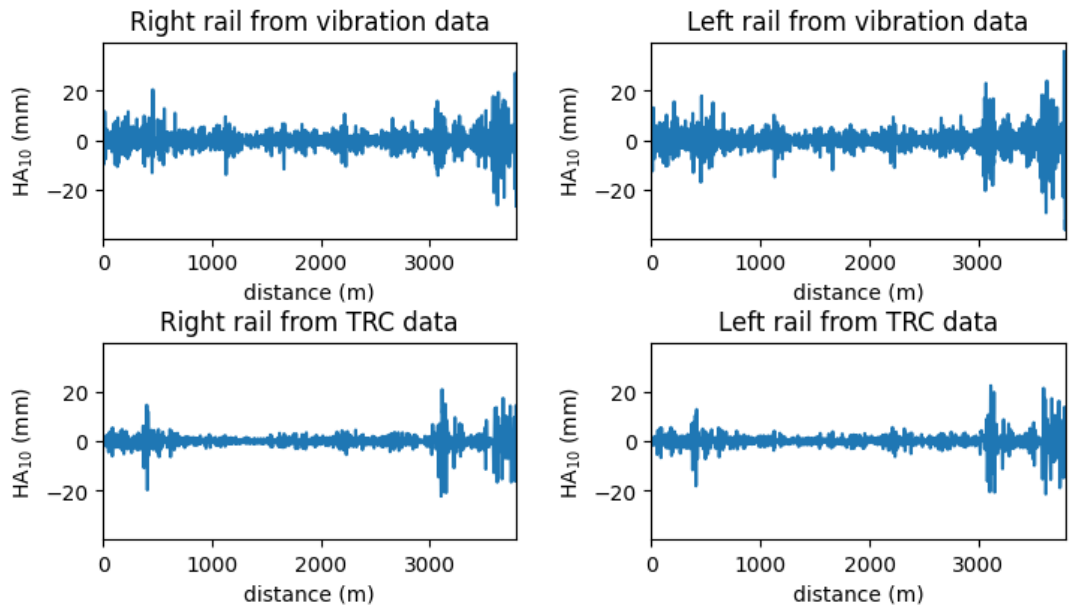

*Figure 17. Comparison of $HA_{10}$.*

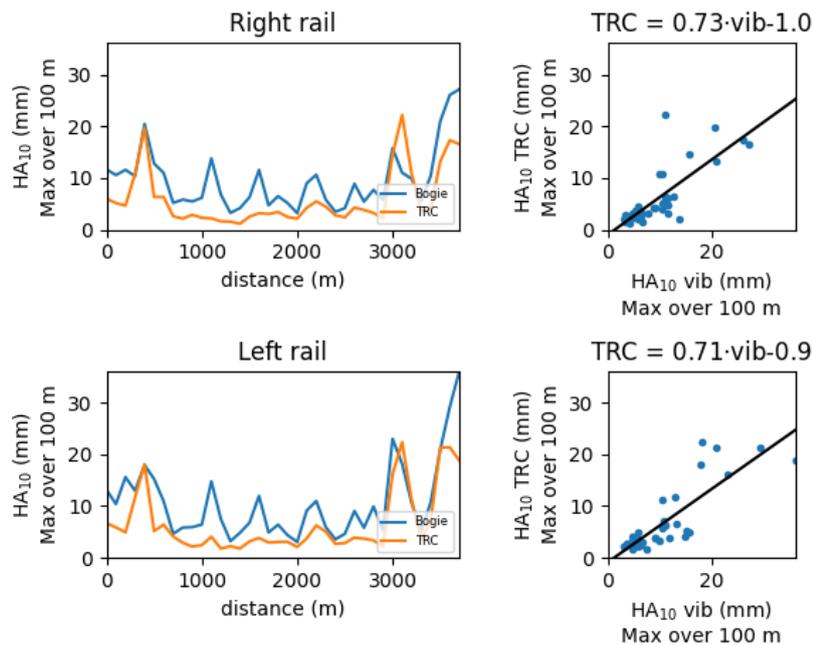

*Figure 18. Comparison of maximum $HA_{10}$ of every 100 meters.*

## 4.5 Frequency analysis of track geometry measurements (bogie measurements)

In this section, an analysis of the frequency domain effects of the vertical/horizontal alignment formulas will be discussed to provide justifications for the choice of filtering parameters required to reconstruct displacement from acceleration measurements.

Using vertical alignment as an example, in the spatial-frequency domain, Eq. (4) can be represented as follows:

$$VA_d(v) = Z(v) - \frac{1}{2}\left(e^{-j2\pi v\left(\frac{d}{2}\right)}Z(v) + e^{j2\pi v\left(\frac{d}{2}\right)}Z(v)\right) \tag{6}$$

where $v$ is spatial frequency. The "transfer function" $H_d(v)$ between the vertical displacement and vertical alignment can thus be written as

$$H_d(v) = \frac{VA_d(v)}{Z(v)} = 1 - \cos(\pi d v) \tag{7}$$

$H_d(v)$ is thus a sinusoidal function in the frequency domain as shown in Figure 19, indicating that the resulting $VA_d$ calculated from Eq. (4) should observe the same trend in frequency, but multipled by the spatial-frequency-domain vertical displacement $Z(v)$. In Figure 20, the spatial PSD of the vertical displacement and the corresponding $VA_{10}$ clearly shows this transfer function relationship between the two parameters, using the data from Eagle Junction to Bowen Hills. It can also be observed from both Figure 19 and Figure 20 that zeros occur at even multiples of the reciprocal of $d$, whereas maximum amplification due to $H_d(v)$ occurs at the odd multiples.

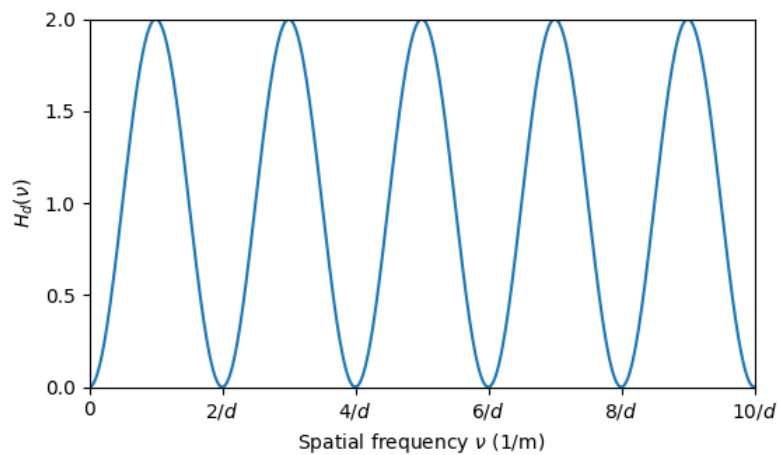

*Figure 19. Transfer function $H_d(v)$ between displacement and vertical/horizontal alignment.*

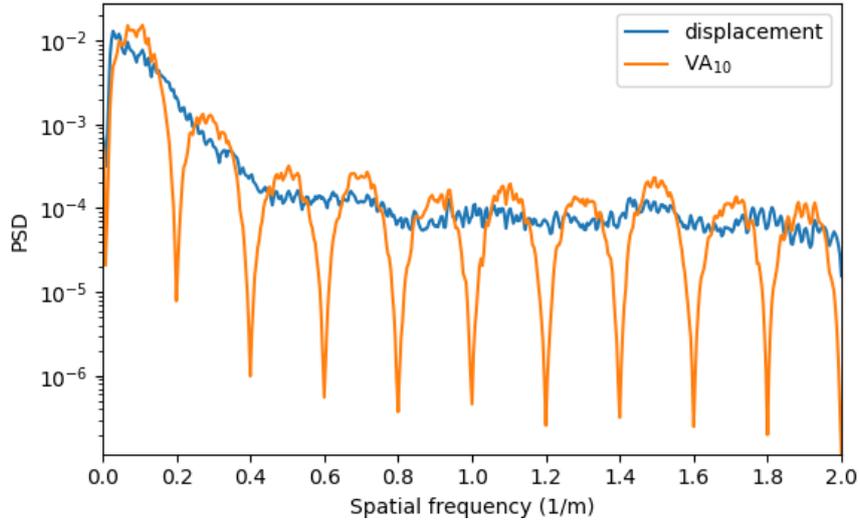

*Figure 20: PSD of vertical displacement and $VA_{10}$.*

In the previous section, 0.3 Hz was used as high-pass cutoff frequency for the bogie measurements when calculating $VA_{10}$. It is now possible to justify this choice by converting the spatial frequency axis $\nu$ into a proper frequency axis $f$, i.e., by multiplying the x-axis of Figure 19 by a series of speed values starting from a "reference low speed" of 3 m/s (~10 km/h), which (as shown in Figure 3, and confirmed by the histogram in Figure 21) represents a lower limit for most of the significant measurements available. The corresponding Alignment "Hz"-transfer functions for $d$ = 10 m are shown in Figure 22 (top). The frequency range below 0.3 Hz is attenuated (with respect to the maximum of the transfer function) for any speed greater than 3 m/s, and significantly reduced for the average speed of ~12 m/s recorded over the 0.25 m segments.

Applying the same consideration for $d$ = 35 m, the cutoff frequency of 0.1 Hz was selected. Figure 22 (bottom) shows the effect of speed on the Alignment transfer function for $d$ = 35 m. The same phenomenon should affect equally $VA_d$ and $HA_d$ since they are calculated in the same way based on vertical and lateral displacements respectively. This frequency analysis of the transfer function effect of the alignment parameter formulas thus allows justifying the choice of the cutoff frequency of the high-pass filter necessary when converting acceleration measurements to displacement, thus addressing a general problem when using acceleration signals for track monitoring.

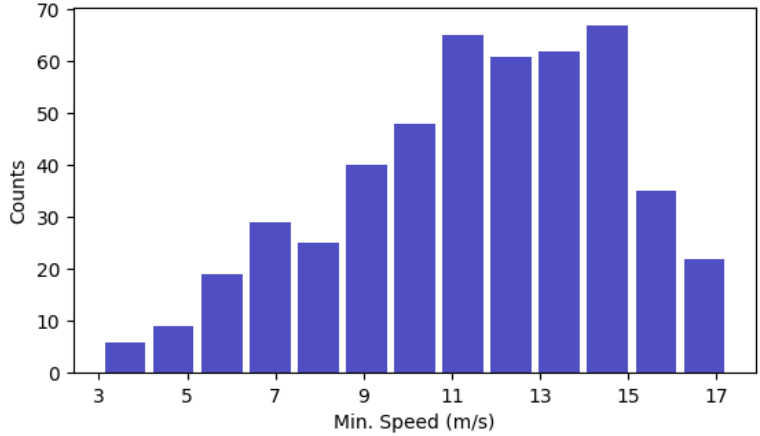

*Figure 21: Histogram of the minimum speed recorded over each 100 m segment.*

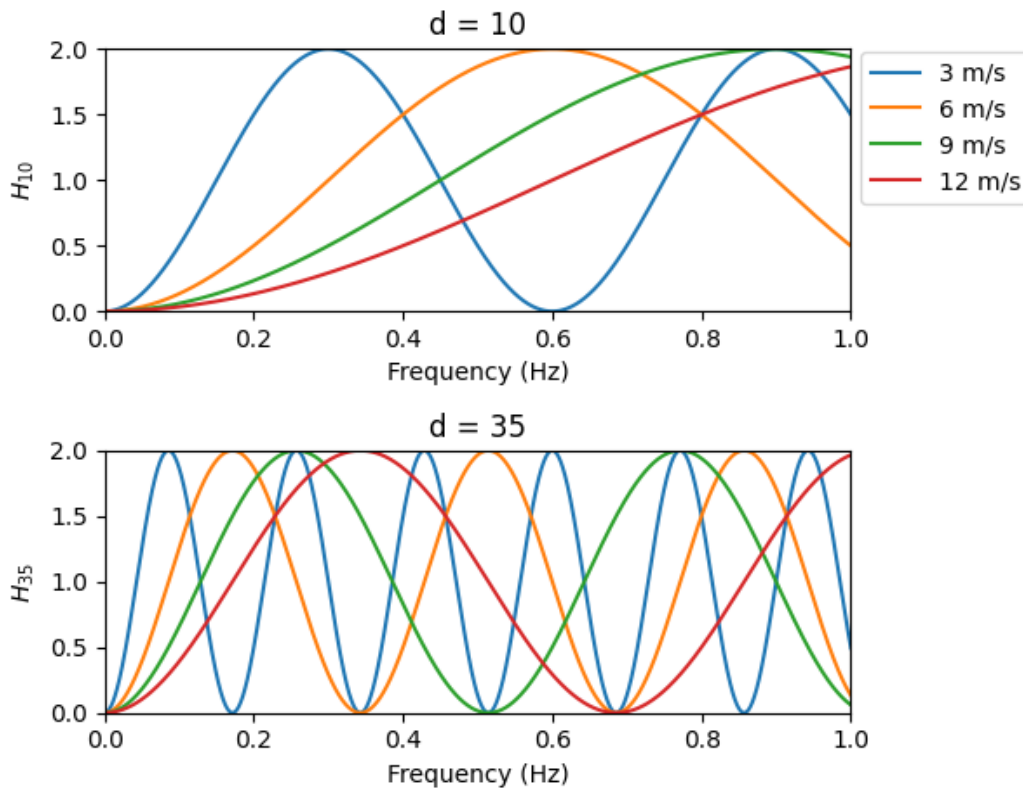

*Figure 22: Displacement-to-alignment transfer function as a function of frequency (Hz) for $d = 10$ (top) and $d = 35$ (bottom) for different speeds.*

## 5. Conclusions and next steps

This paper presents the results of the first test campaign with a prototype onboard track condition monitoring system using affordable MEMS accelerometers. It has shown that key condition indicators (e.g., multiple vertical alignment measurements) can be successfully estimated from the measured acceleration signals, while others show strong correlation with onboard measurements (e.g., horizontal alignment). The bogie sensors were found particularly suited for the application, outperforming axle box and car body installations. The MEMS technology (among the cheapest available) was proven sufficiently sensitive and accurate, thus suggesting a potential for scalability of onboard systems to fleets. A large part of the estimation errors is likely due to signal alignment issues, arising from the fact that a GPS unit was not available for the vibration DAQ and timestamps were not recorded by the TRC system, thus rendering particularly arduous and approximate the comparison of a dataset sampled in time (vibration) and one based on track-distance (TRC). The theoretical frequency analysis of the vertical/horizontal alignment demonstrated in this paper also allows a justification for the choice of the high-pass filter when trying to reconstruct displacement from acceleration signals, which can serve as a future guide for this processing procedure to obtain a reasonable filtered displacement.

These promising results pave the way for a series of future development, mainly in two areas:

1. Refinement of the signal-processing of vibration sensors. This includes considerations on exact sensor locations on the bogie sensors for a more accurate estimation of the profiles of the two tracks.
2. Optimisation of the onboard system, with a lower number of vibration sensors, GPS capabilities, and potentially additional sensor technologies. Regarding the latter, it is to be seen if other options of inertial measurement like gyroscopes are suitable for the estimation of angular parameters such as "super" and "twist".

**Acknowledgements**

This work received support from Queensland Rail (QR) and the Australian Government through the Australian Research Council (ARC) Linkage Project LP200100382. The authors would also like to acknowledge the significant experimental help from QR employees for the organisation of the test campaign, the installation of the onboard system, and the TRC run.

**Appendix**

This section demonstrates the complete procedure to estimate the instantaneous vehicle speed using two bogie vertical measurements on the same side of the bogie, which are expected to have a strong correlation, with a time-varying time-delay $\Delta t[n]$ which is approximated as the wheelbase (2.5 m) divided by the instantaneous vehicle speed $s[n]$ in m/s:

$$\Delta t[n] = 2.5/s[n] \tag{A1}$$

The cross-correlation between the two acceleration signals allows the identification of the most likely delay, and thus the instantaneous speed. This obviously assumes that most of the signal measured by each accelerometer is directly affected by the track profile observed at each instant by the closest wheel:

$$\Delta t[n] = \frac{1}{f_s} \underset{m'}{\mathrm{argmax}} \left( \sum_{m=-N_w/2}^{N_w/2-1} z_{front}[n+m] \, z_{back}[n+m+m'] \right) \quad (A2)$$

with $f_s = 256$ Hz representing the sampling rate in samples/s, $z_{front}[n]$ and $z_{back}[n]$ representing the displacement signals obtained from the front-right and back-right accelerometers on the bogie, and $N_w = 960$ samples used as a moving window to calculate a time-varying delay along the signal record.

The instantaneous speed could then be obtained as $s[n] = 2.5/\Delta t[n]$ and compared to that recorded in the TRC data for horizontal-axis alignment (starting point) by cross correlation.